\documentclass[MSNbibl,nameyear,dvips]{arxstspdf}
\usepackage{flushend}
\usepackage{stfloats}
\usepackage{graphicx}

\volume{26}
\issue{3}
\pubyear{2011}
\firstpage{322}
\lastpage{325}
\doi{10.1214/11-STS352C}
\referstodoi{10.1214/11-STS352}

\begin{document}
\begin{frontmatter}
\vspace*{12pt}
\title{Frasian Inference}
\runtitle{Discussion}
\pdftitle{Discussion of Is Bayes Posterior just Quick and Dirty Confidence? by D.~A.~S.~Fraser}

\begin{aug}
\author[a]{\fnms{Larry} \snm{Wasserman}\corref{}\ead[label=e1]{larry@stat.cmu.edu}}
\runauthor{L. Wasserman}


\address[a]{Larry Wasserman is Professor,
Department of Statistics and
Machine Learning Department,
Carnegie Mellon University,
Pittsburgh, Pennsylvania 15217, USA
\printead{e1}.}

\end{aug}

%
\begin{abstract}
Don Fraser has given an interesting account of
the agreements and disagreements between Bayesian posterior probabilities
and confidence levels.
In this comment
I discuss some cases where the lack of such
agreement is extreme.
I then discuss a few cases where it is possible to
have Bayes procedures with frequentist validity.
Such frequentist-Bayesian---or \textit{Frasian}---methods
deserve more attention.
\end{abstract}

\end{frontmatter}

\section{Introduction}\vspace*{3pt}

Don Fraser has long advocated
the idea that users of Bayesian methods have an obligation
to study the frequentist properties of those methods.
He makes the case quite forcefully
when he states:
``The failure to make true assertions with a promised reliability
can be extreme with the Bayes use of mathematical priors'' and,
more ominously:
\begin{quote}
The claim of a probability status for a~sta\-tement that can fail
to approximate confidence is misrepresentation.
In other areas of science such false claims would be treated
seriously.
\end{quote}

I completely agree with Don and I enjoyed reading his essay
highlighting cases where
approximate confidence does or does not hold.
In this comment
I will mention a few other places where
Bayes methods have poor frequentist coverage.
Then, on a more optimistic note,
I'll discuss a few cases where
Bayes methods do have good frequentist properties.
I'll refer to these methods
as Frasian, both to honor the author and
as a handy way to
refer to methods that meld
frequentist guarantees with Bayesian ideas.

\section{High-Dimensional Models}\vspace*{3pt}

Don's article shows that even in low-dimensional parametric
models,
Bayesian probability statements and confidence statements can
diverge in nontrivial ways.
The situation can be dramatically worse in
high-dimensional and infinite dimensional models.

\subsection*{DKW versus DP}
A simple example concerns estimating the cumulative distribution
function $F$. Let $X_1, \ldots, X_n \sim F$. Let $F_n(x) =
\frac{1}{n}\sum_{i=1}^n I(X_i \leq x)$ be the usual empirical
distribution function. By the famous DKW (Dvo\-retsky--Kiefer--Wolfowitz)
inequality, we know that
\[
\sup_F \mathbb{P}_F \Bigl( \sup_x |F_n(x) - F(x)| > \varepsilon\Bigr) \leq2
e^{-2n\varepsilon^2}.
\]
Hence,
\begin{eqnarray*}
&&(L(x),U(x)) \\
&&\quad\equiv\bigl( \max\{F_n(x) - \varepsilon_n, 0\}, \min\{
F_n(x) + \varepsilon_n, 1\} \bigr)
\end{eqnarray*}
is a valid $1-\alpha$ confidence band, if we set
$\varepsilon_n = \sqrt{\frac{1}{2n}\log(2/\alpha)}$.
(Of course,\vspace*{1pt} narrower bands are possible.)

The usual Bayesian approach is the DP (Dirichlet Process) approach.
Here, $F$ is a given Dirichlet process prior
with mean $F_0$ and concentration parameter $\beta$, $F\sim\operatorname{DP}(F_0,\beta)$.
The posterior
is
$\operatorname{DP}(\overline{F}_n,\beta+n)$
where $\overline{F}_n = \frac{\beta}{\beta+n}F_0 + \frac{n}{\beta+n}F_n$.
Let $(L,U)$ be\vspace*{1pt} a posterior $1-\alpha$ confidence band.
In general, the coverage
$\inf_F \mathbb{P}( L \leq F \leq U)$ is 0.
This is a striking deviation from
frequentist validity.
The frequentist estimator can be recovered by
formally letting $\beta\to0$,
although doing so is to just give up on Bayes.

\subsection*{Normal Means}
Let $Y_i = \theta_i + \frac{1}{\sqrt{n}}\varepsilon_i$, $i=1,2,\ldots,$
where $\varepsilon_1,\varepsilon_2, \ldots$ are $N(0,1)$. This is the standard
Normal means problem and many other problems, such as nonparametric
regression, have been shown to be equivalent to this problem.

Suppose that
$\theta= (\theta_1,\theta_2,\ldots )$ is in the Sobolev ellipsoid
\[
\Theta= \Biggl\{ \theta\dvtx \sum_{j=1}^\infty\theta_i^2 i^{2p} \leq C^2
\Biggr\}.
\]
This corresponds to smooth regression functions in the nonparametric
regression problem.
The minimax rate
is $n^{-2p/(2p+1)}$ and simple shrinkage estimators
achieve this risk.
Zhao (\citeyear{Zha00}) and Shen and Wasserman (\citeyear{SheWas01}) showed that
the priors that yield posterior that achieve the minimax rate
are quite strange and unnatural and are never used in practice.
The obvious prior---Normal on each coordinate---is not minimax unless we allow the prior to put zero mass on $\Theta$.
This hints at the difficulties inherent in melding
Bayes and frequentist ideas in high dimensions.

It gets worse when we look at the type of validity that Don focuses on.
Can we find a prior in this problem such that the $1-\alpha$
posterior regions also have approximate $1-\alpha$ coverage?
To the best of my knowledge, there is no definitive answer.
But the results in Cox (\citeyear{Cox93}) and Freedman (\citeyear{Fre99}) suggest that
the answer is no.

\subsection*{Missing Data and Causal Inference}
Robins and Ritov (\citeyear{RobRit97}) construct an example that is
motivated by missing data problems and causal inference problems. I
refer the reader to their paper for details. But the punch line is
dramatic. The frequentist interval (based on the Horwitz--\break Thompson
estimator) shrinks at rate $O(1/\sqrt{n})$. For a Bayesian region to
have correct coverage, its size will have to shrink no faster than a
logarithmic rate. Hence, there is a drastic loss in efficiency if we
want validity.

\section{Frasian Inference}

Is it possible to force Bayesian methods to have frequentist guarantees?
Don's article shows that the answer can be subtle.
It depends on the structure of the model.
Here I highlight two general techniques
where we can force the Bayesian procedure to have
finite sample frequentist guarantees.

\subsection*{Prediction}
Let $\pi(\theta|Y^n) \propto f(Y^n|\theta) \pi(\theta)$ denote the
posterior where $Y^n = (Y_1,\ldots, Y_n)$. The predictive distribution
for a new observation $Z$ (drawn from the same distribution as $Y^n$)
is $\pi(z|Y^n) = \int f(z|\theta)\pi(\theta|Y^n) \,d\theta$. The usual
Bayesian approach for prediction is to choose a set $B$ such that
$\int_B \pi(y|Y^n) \,dy = 1-\alpha$. Of course, $B$~need not have
frequentist coverage validity.

\begin{figure*}[t!]

\includegraphics{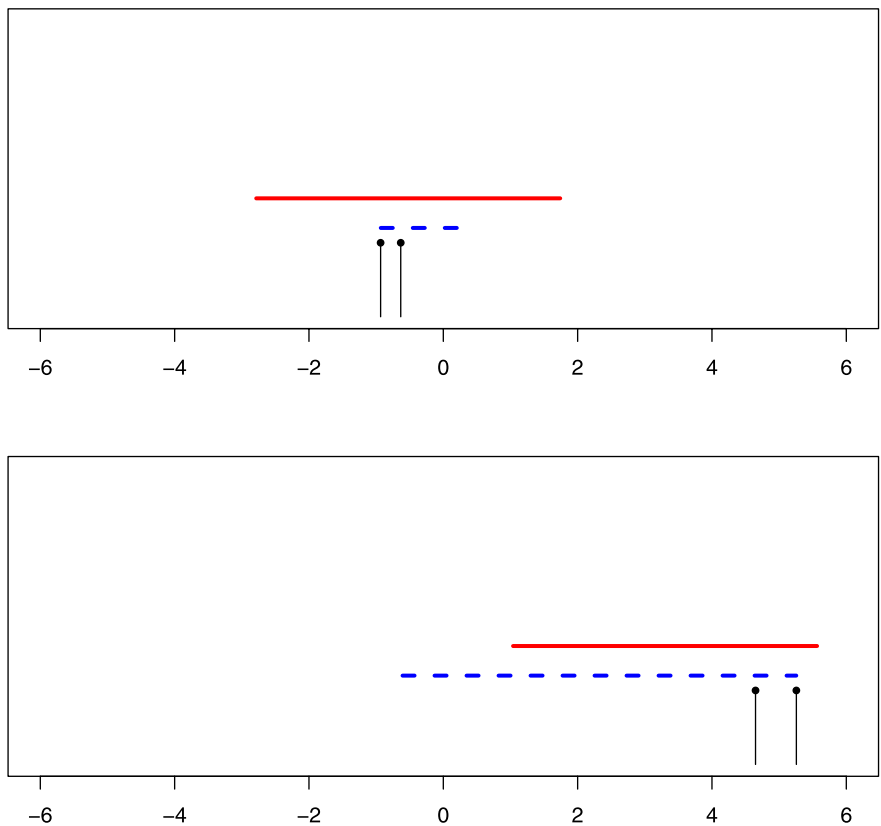}

\caption{In the top plot $Y_1,Y_2 \sim N(0,1)$.
In the bottom plot $Y_1,Y_2 \sim N(5,1)$.
In both cases, the prior is $\theta\sim N(0,1)$.
The two vertical lines show the locations of the two data points.
The dashed horizontal line is the frequentized region.
The solid horizontal line is Bayes predictive region.}\label{fig::plot}
\end{figure*}

But we can adapt the ideas in
Vovk, Gammerman and Shafer (\citeyear{VovGamSha05})
to get a predictive region~$A$
with finite sample frequentist validity.
To construct~$A$, we test the null hypothesis
$H_0\dvtx Z=z$
using the Bayesian predictive density as a test statistic.
We then invert the test to get $A$.
Here are steps in detail:

\begin{enumerate}[1.]
\item[1.] Fix $Z$ at some value $z$.
\begin{longlist}[(a)]
\item[(a)]
Set $Y_{n+1} =z$ and
form the augmented data set
$Y_1,\ldots, Y_n,Y_{n+1}$.
\item[(b)] Compute the predictive density
$\pi(\cdot|Y_1,\ldots,\allowbreak  Y_n,Y_{n+1}) = \int f(\cdot|\theta)\pi(\theta
|Y_1,\ldots, Y_n,Y_{n+1}) \,d\theta$.
\item[(c)] Compute the discrepancy statistics
$D_1,\ldots,\allowbreak D_{n+1}$ where
$D_i\,{=}\,\pi(Y_i |Y_1,\ldots,Y_n,Y_{n+1}), i\,{=}\,1,\ldots,\allowbreak n+1$.
\item[(d)] Compute the $p$-value $p(z)$ for testing $H_0\dvtx\allowbreak Z=z$ by
\[
p(z) = \frac{1}{n+1}\sum_{i=1}^n I(D_i \leq D_{n+1}).
\]
Under $H_0$,
$D_1,\ldots, D_{n+1}$ are exchangeable so this is a valid $p$-value.
\end{longlist}
\item[2.] After computing the $p$-value $p(z)$ for each value~$z$,
invert the test: let
\[
A = \{z\dvtx p(z) \geq\alpha\}.
\]
\end{enumerate}

It follows that
\[
\mathbb{P}(Z \in A) \geq1-\alpha.
\]
This is true no matter what the prior is.
In fact, it is true even if the model is wrong.
Using the Bayesian predictive region as a test statistic is how we let the
prior enter the problem.
A good prior might lead to small prediction regions $A$.
Thus, validity is guaranteed; only efficiency is in question.
Here we are making use of the Bayesian machinery while
maintaining frequentist validity.
I will refer to $A$ as the frequentized region.

Figure \ref{fig::plot} shows a toy example.
The data are $N(\theta,1)$ and the prior is
$\theta\sim N(0,1)$.
To make the effect clear, we use\vadjust{\goodbreak} a tiny sample size of $n=2$
and we use $\alpha= 0.05$.
The top plot shows the case where $\theta=0$ so the prior
is consistent with the truth.
The two vertical lines show the locations of the two data points.
The dashed horizontal line is the frequentized region.
The solid horizontal line is Bayes predictive region.\looseness=-1

The second plot shows an example with $\theta=5$.
Here there is a conflict between the prior and the truth.
The Bayes region is shorter but of course does not have
frequentist validity.
The frequentized region
is longer. This is the compensation for having a bad prior.

\subsection*{Weighted Hypothesis Testing}
Consider testing $m$ null hypotheses $H_{01},\ldots, H_{0m}$ based on
$p$-values $P_1,\ldots, P_m$. The Bonferroni me\-thod takes the rejection
set to be $R = \{ j\dvtx P_j \leq\alpha/m\}$. It is well known that this
procedure controls the error rate in the sense that
%
\begin{equation}\label{eq::guar}
\mathbb{P}(R \cap\mathcal{H}_0 \neq\varnothing) \leq\alpha,
\end{equation}
where
$\mathcal{H}_0 = \{ j\dvtx H_{0j} \mbox{ is true}\}$.

Suppose we have prior information that favors some of these null
hypotheses.\vadjust{\goodbreak}
We could include this~prior information by adopting a Bayesian\break \mbox{analysis}.
But then we lose the frequentist guarantee
given in~(\ref{eq::guar}).
Is there a way to tilt the analysis according to our prior information while
preserving~(\ref{eq::guar})?
The \mbox{answer} is yes.
Simply replace the \mbox{$p$-values}
by weighted $p$-values
$P_j/w_j$ and carry out the Bonferro\-ni~procedure.
As long as the prior\break weights are non-negative and sum to one,
then (\ref{eq::guar}) still holds.
(See Roeder and Wasserman, \citeyear{RoeWas09}, and
Genovese, Roeder and Wasserman, \citeyear{GenRoeWas06}.)
Although not formally a Bayesian~procedure, it does allow us to
have a nugget of Baye\-sianism by including prior weights
while still preserving the frequentist guarantee.

For one-sided testing of Normal means,
the optimal weights are
$w_j = (m/\alpha)\overline{\Phi}(\theta_j/2 + c/\theta_j)$,
where
$\overline{\Phi}$ is the Gaussian survivor function and $c$
is the constant that makes the weights sum to one.
The optimal weights depend on the unknown means
$\theta_j$.
Here is another opportunity to blend frequentist with\break Bayes~by using a prior
on the $\theta_j$'s to optimize the weights.

\section{Conclusion}

Don Fraser has shown that, except in special circumstances,
Bayesian posterior probabilities and frequentist confidence can diverge.
The degree of divergence depends on features
of the model such as nonlinearity.

I have discussed cases where the divergence
can be extreme.
On the other hand, I have also discussed some approaches
for forcing Bayesian methods to have frequentist validity.
But in general, we must be vigilant and pay careful attention to
the sampling properties of procedures.
Don's paper is a useful reminder of the need for that vigilance.


\end{document}